# Multidimensional hybrid Bose-Einstein condensates stabilized by lower-dimensional spin-orbit coupling

Y. V. Kartashov,[1,2] L. Torner,[1,3] M. Modugno,[4,6] E. Ya. Sherman,[5,6]
B. A. Malomed,[7] and V. V. Konotop[8]

[1]*ICFO-Institut de Ciències Fotòniques, The Barcelona Institute of Science and Technology, 08860 Castelldefels (Barcelona), Spain*
[2]*Institute of Spectroscopy, Russian Academy of Sciences, Troitsk, Moscow, 108840, Russia*
[3]*Universitat Politècnica de Catalunya, 08034, Barcelona, Spain*
[4]*Department of Theoretical Physics and History of Science, University of the Basque Country UPV/EHU, 48080 Bilbao, Spain*
[5]*Department of Physical Chemistry, The University of the Basque Country UPV/EHU, 48080 Bilbao, Spain*
[6]*IKERBASQUE Basque Foundation for Science, 48013 Bilbao, Spain*
[7]*Department of Physical Electronics, School of Electrical Engineering, Faculty of Engineering, and Centre for Light-Matter Interaction, Tel Aviv University, 69978 Tel Aviv, Israel*
[8]*Departamento de Física and Centro de Física Teórica e Computacional, Faculdade de Ciências, Universidade de Lisboa, Campo Grande, Edifício C8, Lisboa 1749-016, Portugal*

We show that attractive spinor Bose-Einstein condensates under the action of spin-orbit coupling (SOC) and Zeeman splitting form self-sustained stable two- and three-dimensional (2D and 3D) states in free space, even when SOC acts in a lower-dimensional form. We find that two-dimensional states are stabilized by one-dimensional (1D) SOC in a broad range of chemical potentials, for atom numbers (or norm of the spinor wavefunction) exceeding a threshold value, which strongly depends on the SOC strength and vanishes at a critical point. The zero-threshold point is a boundary between single-peaked and striped states, realizing *hybrids* combining 2D and 1D structural features. In a vicinity of such point, an asymptotic equation describing the bifurcation of the solitons from the linear spectrum is derived and investigated analytically. We show that striped 3D solitary states are as well stabilized by 2D SOC, albeit in a limited range of chemical potentials and norms.

PhySH Subject Headings: Solitons; Superfluids; Mixtures of atomic and/or molecular quantum gases

### I. Introduction and model

Unlike one-dimensional (1D) settings, where stable soliton states exist in diverse systems, stability is a major issue for multidimensional self-trapped localized states [1,2]. In 2D and 3D settings with the ubiquitous cubic self-attraction, instability of fundamental solitons is driven, respectively, by critical and supercritical collapse [3-5], while vortex solutions are subject to azimuthal self-splitting instability [2]. Several stabilization mechanisms for multidimensional solitons have been elaborated theoretically. In a recent landmark advance, stabilization of Bose-Einstein condensates (BECs) against collapse [6,7] by Lee-Huang-Yang (LHY) quantum corrections to the mean-field interactions [8] has been demonstrated experimentally, leading to the creation of fundamental (zero-vorticity) solitary states in the form of "quantum droplets" in BECs with dipole-dipole [9-11] and with contact [12-15] interactions. Regarding "droplets" with embedded vorticity, they are unstable in the former case [16], while LHY-stabilized 3D [17] and 2D [18,19] vortical droplets have been predicted in BECs with contact nonlinearity.

Another possibility for the creation of stable solitons in atomic BECs was predicted in the model of spinor condensates with cubic attraction and spin-orbit coupling (SOC) between the two components [20-22] (see [23] for a similar result for optical solitons and [24] for the stabilization of solitons in BECs by Zeeman lattices). Previously, it was assumed that the cubic self-attraction in the 2D geometry always leads to critical collapse when the norm of the wave function, $U$, exceeds a critical value, $U_{cr}$, while any input decays at $U < U_{cr}$, hence the soliton solution existing at $U = U_{cr}$ is unstable [3,4]. In Refs. [20-22] it was shown that SOC may change the situation at $U < U_{cr}$, creating stable 2D solitons, which represent the otherwise missing ground state in the system. In 3D, the supercritical collapse (which in the presence of SOC was studied in [25]) does not let SOC create the ground state, although 3D soliton solutions stable against small perturbations have been predicted [26].

In this paper, we explore the possibility to create stable 2D and 3D self-trapped states supported by *lower-dimensional* SOC (1D and 2D, respectively). Such settings suggest the existence of new species of multidimensional solitons, which strongly differ from the solitons supported by the full SOC. Their stability is a particularly challenging issue, as reduction of the dimensionality of the support structure makes it harder to secure the stability, cf. the study of fundamental and vortex 2D and 3D solitons supported, respectively, by the 1D and 2D lattice potentials [2,27-29]. The use of lower-dimensional SOC settings offers a crucial advantage for the experimental creation of solitons, as the majority of experimental realizations of SOC were performed in the 1D geometry [30-32], with 2D schemes implemented in a few works [33,34], but not yet in 3D.

The evolution of a spinor BEC in space of dimension $D$ is modelled by coupled Gross-Pitaevskii equations (GPEs) for a spinor wave-function $\mathbf{\Psi} = (\psi_1, \psi_2)^T$ [1,30-32]:

$$i\frac{\partial \mathbf{\Psi}}{\partial t} = -\frac{1}{2}\Delta_D \mathbf{\Psi} - i\sum_{\xi=x,y}\alpha_\xi (\mathbf{h}^{[\xi]} \cdot \boldsymbol{\sigma})\frac{\partial}{\partial \xi}\mathbf{\Psi} + \frac{\Omega}{2}\sigma_z \mathbf{\Psi} - \mathcal{N}\mathbf{\Psi}. \quad (1)$$

Here $\alpha_x, \alpha_y$ are SOC strengths associated with the respective spatial directions, defined by unit-length vectors $\mathbf{h}^{[x]}, \mathbf{h}^{[y]}$, $\boldsymbol{\sigma} = (\sigma_x, \sigma_y, \sigma_z)$ is the vector of Pauli matrices, and $\Omega$ is the Zeeman splitting (ZS), cf. Refs. [22,35,36]. The choice of vectors $\mathbf{h}^{[x]}$ and $\mathbf{h}^{[y]}$ determines the symmetry type of SOC in the system of interest. The nonlinearity is represented by a diagonal $2\times 2$ matrix, $\mathcal{N} = \text{diag}(g_s|\psi_1|^2 + g_c|\psi_2|^2, g_s|\psi_2|^2 + g_c|\psi_1|^2)$, admitting a difference of the cross- ($g_c$) and self- ($g_s$) interactions ($g_s = g_c$ corresponds to Manakov's symmetric system [37]). Localized solutions of Eq. (1) are characterized by the norm, $U = \langle \mathbf{\Psi}|\mathbf{\Psi}\rangle$.

The analysis for 2D and 3D solitons, which are supported, respectively, by the reduced one- and two-dimensional SOC, is presented below in Sections II and III. Results of a systematic numerical investi-gation are combined with analytical findings, which are based on the consideration of the spectrum of the linear version of Eq. (1), as well as on prediction of the stability of the multidimensional solitons produced by the well-known Vakhitov-Kolokolov (VK) criterion. In addition, essential results for 2D solitons are obtained by means of the variational approximation (VA). Results are summarized in Section IV, which also outlines possibility of their experimental implementation.

## II. 2D system with 1D spin-orbit coupling

First, we consider the 2D spinor BEC, with $\Delta_2 = \partial^2/\partial x^2 + \partial^2/\partial y^2$ and 1D SOC defined by $\alpha_y = 0$, $\alpha_x \equiv \alpha$, and $\mathbf{h}^{[x]} = \hat{\mathbf{x}}$ (the unit vector along the $x$- axis) in Eq. (1) corresponding to the SOC Hamiltonian, $H_{\mathrm{soc}} = \alpha k_x \sigma_x$. In this system, SOC may be gauged away by a canonical transformation, $\tilde{\Psi} \to S\Psi$, corresponding to a position-dependent spin rotation $S = \exp(-i\alpha\sigma_x x)$, resulting in a non-Abelian 1D Zeeman lattice field, in the form of $\tilde{\Omega}(x) = (\Omega/2)[\sigma_z \cos(2\alpha x) + \sigma_y \sin(2\alpha x)]$ [36]. Then, one may expect that such a Zeeman lattice stabilizes 2D solitons [28,29]. However, it is more relevant to start the analysis with identifying the linear spectrum of Eq. (1) for the chemical potential $\mu$ that can be obtained by substitution $\Psi = \mathbf{C} e^{ik_x x + ik_y y - i\mu t}$, where $\mathbf{C}$ is constant spinor and $k_{x,y}$ are wavenumbers of the respective excitation, into the linearized version of Eq. (1). A straightforward calculation yields the spectrum which consists of two branches,

$$\mu_\pm = (k_x^2 + k_y^2)/2 \pm (\alpha^2 k_x^2 + \Omega^2/4)^{1/2}. \quad (2)$$

The critical nature of the 2D collapse in the system with $\alpha = 0$ (without SOC) is underlain by the fact that both the kinetic and nonlinear-interaction energies scale as $k_{x,y}^2$, while the total norm takes a single value $U = U_{\mathrm{cr}}$ for all solitons. The breaking of the scaling (conformal) invariance by SOC (with $\alpha \neq 0$) and lifting the norm's degeneracy may create a stability range for 2D solitons at $U < U_{\mathrm{cr}}$ [20,21].

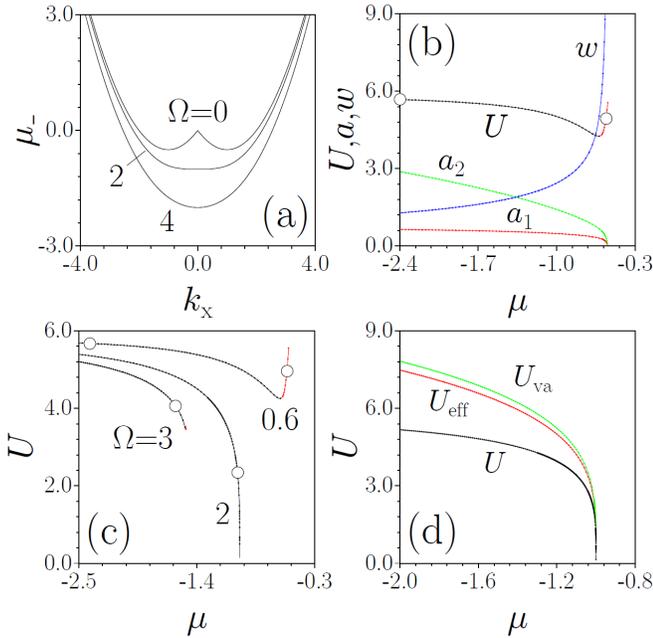

Fig. 1. (a) Lower branch $\mu_-(k_x)$ of 2D linear spectrum (2) for $k_y = 0$, $\alpha = 1$ and different values of the ZS strength, $\Omega$. (b) Norm $U$, size $w$, and amplitudes $a_{1,2}$ of the components of 2D solitons vs. $\mu$ at $\Omega = 0.6$, $\alpha = 1$. (c) The norm vs. $\mu$ for different values of $\Omega$ at $\alpha = 1$. Circles in (b,c) correspond to solitons depicted in Fig. 2. In (c), as well as in Fig. 4(a) below, stable and unstable branches are black and red, respectively. (d) The $U(\mu)$ dependence for 2D solitons in a vicinity of the transition point, $\Omega = 2\alpha^2$, as produced by the numerical solution of Eq. (1) ($U$), effective asymptotic equation (3) ($U_{\mathrm{eff}}$), and by the respective variational approximation ($U_{\mathrm{va}}$). In (b)-(d), $g_s = g_c = 1$.

Under the action of the self-attraction, solitons that have the form of $\Psi = e^{-i\mu t}\mathbf{W}(x, y)$, with complex spinor $\mathbf{W} = (w_1(x,y), w_2(x,y))^T$, bifurcate upon variation of chemical potential $\mu$ from the minimum of the lower branch of the linear dispersion relation given by Eq. (2), $\mu_-(k_x, k_y)$, which is attained at $k_y = 0$. Dependence $\mu_-(k_x, 0)$ [Fig. 1(a)] reveals that, for $\Omega > 2\alpha^2$, the minimum, $\mu_-^{\min} = -\Omega/2$, is attained at $k_x^{\min} = 0$. On the other hand, at $\Omega < 2\alpha^2$ there are two equal minima, $\mu_-^{\min} = -\alpha^2/2 - \Omega^2/8\alpha^2$, at $k_x^{\min} = \pm(\alpha^2 - \Omega^2/4\alpha^2)^{1/2}$, hence solitons bifurcate from these points, featuring a striped structure, determined by the respective scale in the $x$ direction, $2\pi/|k_x^{\min}|$. Thus, the transition between cases $\Omega \gtrless 2\alpha^2$ leads to a drastic change in the solitons' shapes.

We first produce soliton families for the Manakov-like system, with $g_s = g_c = 1$. Under the action of SOC, two spinor components of such solitons feature different symmetries and amplitudes, $a_{1,2} \equiv \max|w_{1,2}(x,y)|$. At $\Omega > 0$, only the family with a dominating second component, $a_2 > a_1$, may be stable, therefore we address this case (spin-flipped states, with $a_1 > a_2$, exist too, but numerical results readily confirm their instability, as they realize a maximum of the ZS energy, instead of the minimum). The norm, effective size $w = 2[U^{-1}\langle\Psi|(x^2 + y^2)|\Psi\rangle]^{1/2}$, and component amplitudes $a_{1,2}$ of 2D striped solitons, found at $\Omega < 2\alpha^2$, are shown in Fig. 1(b) as functions of $\mu$. Amplitudes $a_{1,2}$ vanish, and width $w$ diverges, at $\mu \to \mu_-^{\min}$. The striped solitons with relatively small amplitudes feature strong modulation along the $x$-axis [Figs. 2(a$_{1,2}$)]. For large $|\mu|$, the modulation nearly vanishes when the soliton size becomes smaller than the modulation period [Figs. 2(b$_{1,2}$)]. These 2D solitons, featuring the dipole structure in the first component, are drastically different from the self-trapped 2D states ("semi-vortices" and "mixed modes" [20,21]) supported by the full 2D SOC. In addition to the different symmetry, the latter states never show striped patterns. In a sense, the 2D solitons displayed in Fig. 2 are *hybrids*, which combine the 2D stability with structural features resembling those found in 1D solitons, cf. Refs. [23,24].

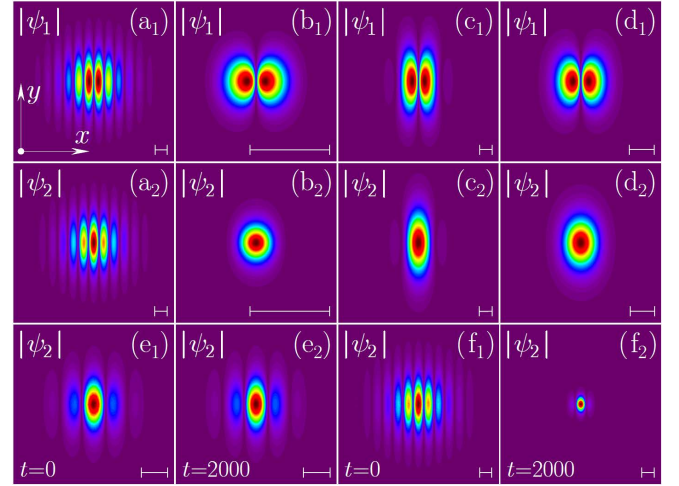

Fig. 2. Profiles $|\psi_{1,2}|$ of 2D solitons for (a) $\mu = -0.56$ and (b) $\mu = -2.40$ at $\Omega = 0.6$, (c) $\mu = -1.02$ at $\Omega = 2.0$, and (d) $\mu = -1.6$ at $\Omega = 3.0$. The evolution of a stable soliton with $\mu = -0.66$ and decay of an unstable one with $\mu = -0.56$ is displayed in (e) and (f), respectively. In all cases, $\alpha = 1$, $g_s = g_c = 1$. Different panels are plotted on appropriate scales, with the horizontal bar corresponding to $x = 4$.

Dependence $U(\mu)$ displayed in Fig. 1(b) is non-monotonous. Also, in contrast to the solitons supported by the full SOC [21,22,25], but similar to 1D states maintained by SOC with the Zeeman lattice [24], the present 2D solitons exist above a threshold value of the norm, $U_{\text{th}} \equiv \min U(\mu)$, which vanishes solely at $\Omega = 2\alpha^2$, being a non-monotonous function of $\Omega$ and $\alpha$. This is illustrated by Fig. 1(c), which compares $U(\mu)$ dependencies for the solitons with the single-peak $(\Omega > 2\alpha^2)$ and striped $(\Omega < 2\alpha^2)$ structures. The transition between these species occurs via shapes strongly elongated along $y$ [Fig. 2(c)], clearly indicating hybrid nature of such states. An example of a single-peaked soliton is presented in Fig. 2(d). In all cases, irrespective of the ratio of $\Omega$ and $\alpha^2$, at $\mu \to -\infty$ the norm approaches that of the *Townes soliton* [38], $U_{\text{Townes}} \approx 5.85$, which is the single possible value for (unstable) 2D solitons in the absence of SOC [3,4].

To explain the vanishing of $U_{\text{th}}$ at the point of transition between the 2D solitons with single-peak and striped structures, $\Omega = 2\alpha^2$, we note that, at this point, the expansion of dispersion relation (2) near the origin ($|k_x| \ll \alpha$) yields $\mu_- = -\alpha^2 + k_y^2/2 + k_x^4/8\alpha^2$. For small deviation of the chemical potential from the minimum, $\mu_-^{\min} = -\alpha^2$, i.e. for $-\Delta\mu \equiv \mu_-^{\min} - \mu \ll |\mu_-^{\min}|$, we explore the bifurcation of solitons from the lower branch of the linear spectrum, i.e., from the state $(0,1)^{\text{T}}$. Accordingly, we look for a stationary solution in the form of $(\psi_1, \psi_2)^{\text{T}} = (0,1)^{\text{T}} \phi(x,y) e^{-i\Delta\mu t} + \mathcal{O}[(-\Delta\mu/\alpha^2)^{3/2}]$, where $\phi(x,y)$ is the slowly varying amplitude. Applying the multiple-scale expansion (see Appendix A for the details of derivation), we find that the amplitude solves a stationary equation,

$$\Delta\mu\phi = -(1/2)\phi_{yy} + (1/4)\phi_{\tilde{x}\tilde{x}\tilde{x}\tilde{x}} - |\phi|^2 \phi, \quad (3)$$

which is written in variables $(\tilde{x}, y) = (2^{1/4}\alpha^{1/2}x, y)$ (cf. Ref. [39]). This equation gives rise to an exact scaling relation, $U(\Delta\mu) \sim (-\Delta\mu)^{1/4}$, which satisfies the necessary stability condition given by the VK criterion, $\partial U/\partial \mu < 0$ [3,4,40]. Soliton solutions of Eq. (3) can be predicted by means of VA [41] based on the Gaussian ansatz with norm $U$:

$$\phi = (U/\pi)^{1/2}(ab)^{1/4}\exp(-a\tilde{x}^2/2 - by^2/2). \quad (4)$$

The VA yields strongly anisotropic relations for parameters of the ansatz: $a = U^2/6\pi^2$, $b = 3a^2/2 \ll a$, and $U = 2\pi(-12\Delta\mu/5)^{1/4}$. The comparison of the $U(\mu)$ dependence produced by Eq. (1) with that obtained from Eq. (3) and its VA counterpart is shown in Fig. 1(d), revealing good agreement at $\mu \to \mu_-^{\min}$.

Dependencies of the threshold norm $U_{\text{th}}$ on $\Omega$ and $\alpha$, as obtained from the numerical solution of Eq. (1), are presented in Fig. 3 (see the curves with solid dots corresponding to $g_s = g_c = 1$). Note that $U_{\text{th}} \to U_{\text{Townes}}$ in both limits of $\Omega \to 0$ (when SOC can be gauged away from Eq. (1) with the Manakov nonlinearity [35,36]) and $\Omega \to \infty$ [making the $\psi_1$ component vanishingly small and reducing Eq. (1) to the single GPE for $\psi_2$].

The stability of the solitons was tested by simulations of Eq. (1) with inputs including random perturbations with relative amplitude $\simeq 1\%$, up to $t > 10^4$. In all cases, the stability exactly follows the VK criterion, the branches with $\partial U/\partial \mu < 0$ and $\partial U/\partial \mu > 0$ [the black and red ones in Figs. 1(b,c)] being stable and unstable, respectively. Thus, 1D SOC stabilizes almost the entire soliton family, except for small segments with $\partial U/\partial \mu > 0$ at $\mu \to \mu_-^{\min}$. It is very plausible that the stable soliton realizes, for given norm, the ground state of the 2D system, although rigorous proof of this conjecture requires additional analysis. Stable propagation of a perturbed 2D soliton is displayed in Fig. 2(e). Unstable broad solitons transform into much narrower stable ones with the same norm, as shown in Fig. 2(f).

Our results remain valid for non-Manakov nonlinearity, with $g_s \ne g_c$, illustrating robustness of the setting. For instance, $U_{\text{th}}(\Omega)$ and $U_{\text{th}}(\alpha)$ dependencies for $g_s = 1.2$ and $g_c = 0.8$, displayed in Fig. 3 by curves with open dots (as well as in their counterparts with $g_s < g_c$), are close to their counterparts obtained for $g_s = g_c = 1$, and they also show vanishing of threshold at $\Omega = 2\alpha^2$.

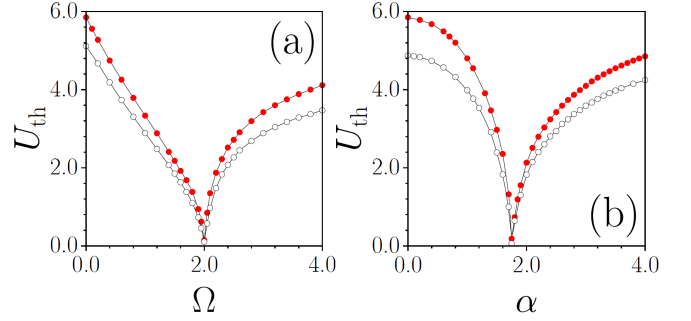

Fig. 3. The threshold norm, above which 2D solitons with the single-peak $(\Omega > 2\alpha^2)$ and striped $(\Omega < 2\alpha^2)$ structure exist, vs. $\Omega$, at $\alpha = 1$ (a), and vs. $\alpha$, at $\Omega = 6$ (b). Curves with solid and open dots correspond, severally, to $g_s = g_c = 1$ and $g_s = 1.2$, $g_c = 0.8$.

### III. 3D system with 2D Rashba spin-orbit coupling

The 3D version of Eq. (1), where $\Delta_3 = \partial^2/\partial x^2 + \partial^2/\partial y^2 + \partial^2/\partial z^2$, is taken with the 2D Rashba SOC, i.e., $\alpha_y = \alpha_x \equiv \alpha$ and $\mathbf{h}^{[x]} = \hat{\mathbf{y}}$, $\mathbf{h}^{[y]} = -\hat{\mathbf{x}}$, corresponding to Hamiltonian $H_{\text{soc}} = \alpha(\hat{k}_x\sigma_y - \hat{k}_y\sigma_x)$, where $\hat{k}_{x,y}$ are the components of the momentum operator. The respective linear dispersion relation for in-plane radial component of the wave vector, $k_r \equiv (k_x^2 + k_y^2)^{1/2}$, and the orthogonal component, $k_z$, is

$$\mu_\pm = (k_r^2 + k_z^2)/2 \pm (\alpha^2 k_r^2 + \Omega^2/4)^{1/2}. \quad (5)$$

The expected single-peak and striped 3D solitons again bifurcate from minima of the $\mu_-$ branch in Eq. (5). Solitons have the form of $\mathbf{\Psi} = (w_1(r,z), w_2(r,z)e^{i\phi})^{\text{T}} \exp(-i\mu t + im\phi)$, where $w_{1,2}(r,z)$ are stationary spinor components, integer $m$ is a topological charge, and $(r,\phi)$ are the polar coordinates in the $(x,y)$ plane. The difference between vorticities of the two components of $\mathbf{\Psi}$, $m$ and $m+1$, is imposed by the 2D SOC. Here we address the modes with $m = -1$, i.e., zero vorticity in the dominant second component, which minimizes the total energy, thus having the best chance to be stable.

The families of 3D solitons at $g_{s,c} = 1$ are displayed in Fig. 4(a). Their amplitude vanishes, and size $w = 2[U^{-1}\langle\mathbf{\Psi}|(r^2+z^2)|\mathbf{\Psi}\rangle]^{1/2}$ diverges, at $\mu \to \mu_-^{\min}$, like in the 2D case. However, their norm vanishes at $\mu \to -\infty$, i.e., 3D solitons have *no existence threshold*. At $\Omega < 2\alpha^2$ the striped structure manifests itself in concentric amplitude-phase modulation in the $(x,y)$ plane [Fig. 5(a)], while the soliton is elongated along the $z$-axis. The interplay of the vorticity in the $\psi_1$ component and striped radial modulation builds a complex phase distribution, which makes the stable 3D solitons radically different from those supported by the full 3D SOC, cf. Ref. [26]. Similar to what is said above about the 2D solitons, the present modes may be considered as *hybrids* combining 3D and 1D features. With the increase of $|\mu|$, they become more localized and the radial modulation gradually disappears [Fig. 5(b)].

As seen in Fig. 4(a), the 3D system with 2D SOC produces striped solitons with a non-monotonous dependence $U(\mu)$, which includes a VK-stable segment with $\partial U/\partial \mu < 0$. At fixed SOC strength $\alpha$, increase of ZS strength $\Omega$ leads to growing inflection of the initially monotonous curve $U(\mu)$. The domain with $\partial U/\partial \mu < 0$ appears at sufficiently large $\Omega$, albeit still for $\Omega < 2\alpha^2$. The VK-stable interval, $U_{\min} < U < U_{\max}$, expands with the increase of $\Omega$, as shown in Fig. 4(b), until the critical point $\Omega = 2\alpha^2$ is attained [see the dashed vertical line in Fig. 4(b)], at which the striped structure is replaced by the single-peak one, and $U(\mu)$ dependence abruptly changes, again becoming monotonous [Fig. 4(a)]. Thus, in contrast to the 2D case, the norm of the 3D solitons does not vanish at $\Omega = 2\alpha^2$, where striped shape is replaced by a single-peak one, while, as in the 2D case, the stability exactly follows the VK criterion, $\partial U/\partial \mu < 0$, even if only the striped solitons are stable in 3D. Similar results are obtained for $g_s \neq g_c$.

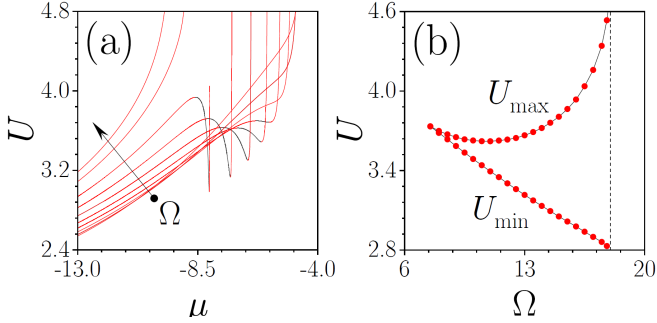

Fig. 4. (a) Norm $U$ of 3D solitons vs. $\mu$ for $\Omega = 2,\ldots,20$, increasing with step $\Delta\Omega = 2$, as shown by the arrow. (b) The stability interval $U_{\min} < U < U_{\max}$ vs. $\Omega$. The dashed vertical line corresponds to $\Omega = 2\alpha^2$. In all cases $\alpha = 3$, $g_s = g_c = 1$.

Unstable 3D solitons are destroyed by the fast collapse, as shown in the top row of Fig. 6. These high-amplitude solitons belong to segments of the $U(\mu)$ curves with $\partial U/\partial \mu > 0$, to the left of the stability domain. The evolution of a stable 3D soliton, belonging to the $\partial U/\partial \mu < 0$ branch, is displayed in the bottom row of Fig. 6. In that case, initial perturbations excite only weak oscillations of the soliton's amplitude. Strong compression of 3D solitons, which are stable against small perturbations, may initiate their supercritical collapse, therefore, they are metastable states, separated by an $\Omega$- and $\alpha$- dependent barrier from the collapse [42].

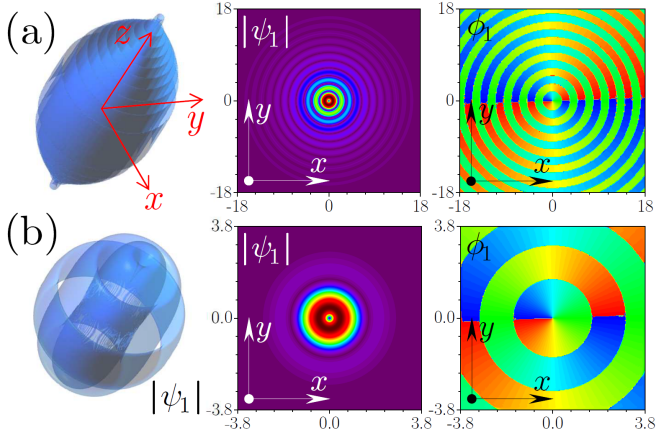

Fig. 5. Isosurfaces $|\psi_1| = 0.001$, showing the shape of 3D solitons, and the respective amplitude and phase profiles in the $z = 0$ plane (the left, central, and right columns, respectively) for (a) $\mu = -6.85$ and (b) $\mu = -7.40$, at $\alpha = 3$, $\Omega = 13$.

In a recent experiment [43] with one-dimensional SOC, the ratio $\Omega/\alpha^2$ has been modified from $0$ to $2.6$, which, according to the above results, allows the existence of stable 2D single-peak and stripe soliton states with the $\alpha$- dependent modulation length $\simeq 3$ $\mu$m. The strength of the nonlinear interaction in 2D, written in physical units, is $g = 4\pi\hbar^2 a_s / M a_z$, where $M$ is the atomic mass, $a_s$ the scattering length (the self-attraction corresponds to $a_s < 0$), and $a_z$ the $z$-axis confinement length. For typical values of $a_s$ and $a_z$ this yields the Townes norm corresponding to $\sim 10^3$ atoms. Note that for two-dimensional SOC [33], the ratio $\Omega/\alpha^2$ is limited by $0.2$, and for maintaining a necessary suppression effect of SOC on the 3D collapse one has to limit the number of atoms to $\leq \hbar^2/4\pi M\alpha a_s \sim 10^3$.

### IV. Conclusions

In summary, we have shown that even a lower-dimensional spin-orbit-coupling than that of the embedding space can stabilize soliton states in spinor BECs with intrinsic attraction, when it is combined with Zeeman splitting. On physical grounds, the stabilization arises from the nonparabolic dispersion of the system and, importantly, holds in a broad parameter region, which complies with the Vakhitov-Kolokolov criterion. We also found that the multidimensional states feature hybridization of 1D and 2D/3D structural properties. The results are important for the creation of stable 2D and 3D solitons in BECs, as spin-orbit coupling is currently experimentally available solely in low-dimensional forms.

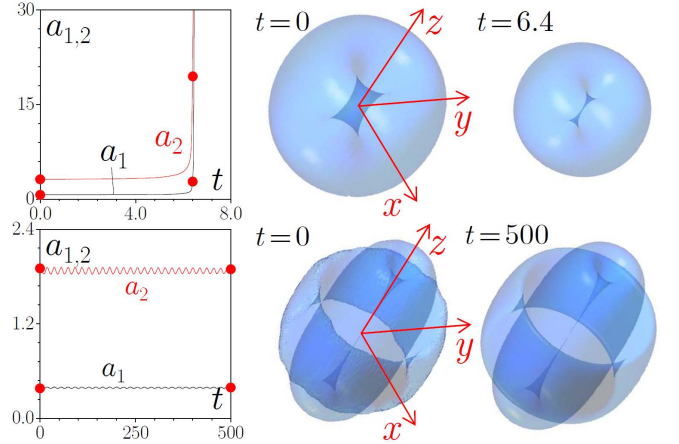

Fig. 6. Top: The collapse of an unstable 3D soliton with $\mu = -8.2$ is shown by the time dependence of amplitudes of its components, and by isosurfaces of the vorticity-carrying one, $|\psi_1| = 0.3$, at two moments of time corresponding to the red dots in the $a_{1,2}(t)$ plot. Bottom: stable evolution of a perturbed 3D soliton with $\mu = -7.4$, illustrated by the respective $a_{1,2}(t)$ dependencies and isosurfaces $|\psi_1| = 0.01$. In both cases $\alpha = 3$, $\Omega = 13$.

### Acknowledgements

L.T. and Y.V.K. acknowledge support from the Severo Ochoa program (SEV-2015-0522) of the Government of Spain, Fundacio Cellex, Fundació Mir-Puig, Generalitat de Catalunya, and CERCA. B.A.M. is supported, in part, by the Israel Science Foundation through grant No. 1287/17. This author appreciates hospitality of ICFO. V.V.K. acknowledges support from the FCT (Portugal) Grant No. UID/FIS/00618/2019. M.M. and E.Y.S. acknowledge support by the Spanish Ministry of Science and the European Regional Development Fund through PGC2018-101355-B-I00 (MCIU/ AEI/FEDER, UE) and the Basque Government through Grant No. IT986-16. E.Y.S. is grateful to M. Glazov and Sh. Mardonov for valuable comments.

### Appendix A: On the small-amplitude limit for the 2D BEC with 1D spin-orbit coupling

Consider a 2D GPE (1) with $\alpha_y = 0$, $\alpha_x \equiv \alpha$, and $\mathbf{h}^{[x]} = \hat{\mathbf{x}}$, as defined in Sec. II. Assuming Manakov's symmetry, we rewrite it for $\tilde{\boldsymbol{\Psi}} = e^{i\alpha^2 t/2}\boldsymbol{\Psi}$ in the form

$$i\frac{\partial\tilde{\boldsymbol{\Psi}}}{\partial t} = \frac{1}{2}P^2\tilde{\boldsymbol{\Psi}} + \frac{\Omega}{2}\sigma_z\tilde{\boldsymbol{\Psi}} - (\tilde{\boldsymbol{\Psi}}^\dagger\tilde{\boldsymbol{\Psi}})\tilde{\boldsymbol{\Psi}}, \tag{A1}$$

where $P = -i(\partial_x + \alpha\sigma_x, \partial_y)$. In order to reduce the number of parameters here we also rescale the coordinates $\mathbf{R} \equiv (X,Y) = \Omega^{1/2}\mathbf{r}$ and introduce $\tilde{\alpha} = \Omega^{-1/2}\alpha$. Looking for a stationary solution of (A1) in the form $\tilde{\boldsymbol{\Psi}} = e^{-i\mu\Omega t}\boldsymbol{\psi}$, where $\boldsymbol{\psi}$ depends on the coordinates only, we obtain

$$\tilde{\mu}\boldsymbol{\psi} = H\boldsymbol{\psi} - (\boldsymbol{\psi}^{\dagger}\boldsymbol{\psi})\boldsymbol{\psi}, \tag{A2}$$

where $\tilde{\mu} = \Omega\mu$ and

$$H = \frac{1}{2}(-i\nabla_{\mathbf{R}} + \hat{\mathbf{x}}\tilde{\alpha}\sigma_x)^2 + \frac{1}{2}\sigma_z. \tag{A3}$$

First, we consider the linear limit of (A2), that is:

$$H\boldsymbol{\psi}_0 = \mu_0\boldsymbol{\psi}_0. \tag{A4}$$

Its orthonormal eigenfunctions $\boldsymbol{\psi}_{0\mathbf{K}}^{\pm}(\mathbf{R})$ and corresponding eigenvalues $\mu_0(\mathbf{K})$ are given by (further $K^2 = K_x^2 + K_y^2$):

$$\boldsymbol{\psi}_{0\mathbf{K}}^{\pm} = [2C(C\mp 1)]^{-1/2} e^{i\mathbf{K}\cdot\mathbf{R}} \begin{pmatrix} \pm 2\tilde{\alpha}K_x \\ C\mp 1 \end{pmatrix}, \tag{A5}$$

where $C = (1 + 4\tilde{\alpha}^2 K_x^2)^{1/2}$, and $\mu_0(\mathbf{K}) = \mu_{\pm}(\mathbf{K})$ where

$$\mu_{\pm} = \frac{1}{2}(K^2 + \tilde{\alpha}^2) \pm \frac{1}{2}(1 + 4\tilde{\alpha}^2 K_x^2)^{1/2} \tag{A6}$$

[the last expression (A6) being the dispersion relation (2) in the dimensionless units]. For the lower dispersion curve the transition from the normal to the stripe phase occurs at $\tilde{\alpha} = \tilde{\alpha}_0 = 2^{-1/2}$. At this value of $\tilde{\alpha}$, $K_x = 0$ is the undulation point of $\mu_{\pm}$ with

$$\left.\frac{\partial\mu_-}{\partial K_x}\right|_{\mathbf{K}=\mathbf{0}} = \left.\frac{\partial^2\mu_-}{\partial K_x^2}\right|_{\mathbf{K}=\mathbf{0}} = \left.\frac{\partial^3\mu_-}{\partial K_x^3}\right|_{\mathbf{K}=\mathbf{0}} = 0. \tag{A7}$$

Since spatially localized nonlinear modes (solitons) bifurcate from the lower branch of the spectrum, $\mu_-$, and we are interested in the special case of $\alpha = \tilde{\alpha}_0$, we conclude that the bifurcation point corresponds to the eigenstate at $\mathbf{K} = \mathbf{0}$ denoted as

$$\boldsymbol{\psi}_0^- = (0,1)^{\mathrm{T}}. \tag{A8}$$

It will be convenient to introduce also a notation for the respective lowest state of the upper band $\mu_+$:

$$\boldsymbol{\psi}_0^+ = (1,0)^{\mathrm{T}}. \tag{A9}$$

Bifurcation of nonlinear modes can be described using multiple-scale expansion. To this end we introduce the slow variables $x_j = \epsilon^{j/2} X$, $y_j = \epsilon^j Y$ (here $j = 0, 1, \ldots$), with $\epsilon \to 0$. The corresponding representation for the linear Hamiltonian at arbitrary $\tilde{\alpha}$ value reads:

$$H = H_0 + \epsilon^{1/2} H_1 + \epsilon H_2 + \epsilon^{3/2} H_3 + \cdots, \tag{A10}$$

where

$$H_0 = -\frac{1}{2}\frac{\partial^2}{\partial x_0^2} - \frac{1}{2}\frac{\partial^2}{\partial y_0^2} - i\tilde{\alpha}\sigma_x\frac{\partial}{\partial x_0} + \frac{\sigma_z}{2} + \frac{\tilde{\alpha}^2}{2},$$

$$H_1 = -\frac{\partial^2}{\partial x_0 \partial x_1} - i\tilde{\alpha}\sigma_x\frac{\partial}{\partial x_1},$$

$$H_2 = -\frac{\partial^2}{\partial x_0 \partial x_2} - \frac{1}{2}\frac{\partial^2}{\partial x_1^2} - \frac{\partial^2}{\partial y_0 \partial y_1} - i\tilde{\alpha}\sigma_x\frac{\partial}{\partial x_2},$$

$$H_3 = -\frac{\partial^2}{\partial x_0 \partial x_3} - \frac{\partial^2}{\partial x_1 \partial x_2} - i\tilde{\alpha}\sigma_x\frac{\partial}{\partial x_3},$$

$$H_4 = -\frac{\partial^2}{\partial x_0 \partial x_4} - \frac{1}{2}\frac{\partial^2}{\partial x_2^2} - \frac{\partial^2}{\partial x_1 \partial x_3} - \frac{\partial^2}{\partial y_0 \partial y_2}$$
$$-\frac{1}{2}\frac{\partial^2}{\partial y_1^2} - i\tilde{\alpha}\sigma_x\frac{\partial}{\partial x_4}.$$

Next we use the expansions for the spinor wavefunction

$$\boldsymbol{\psi} = \epsilon A(x_1; y_1)\boldsymbol{\psi}_0^- + \epsilon^{3/2}\boldsymbol{\psi}_1 + \epsilon^2\boldsymbol{\psi}_2 + \cdots, \tag{A11}$$

and for the chemical potential

$$\tilde{\mu} = \mu_0^- + \epsilon^{1/2}\mu_1 + \epsilon\mu_2 + \cdots, \tag{A12}$$

where we use the notation $\mu_0^{\pm} = \tilde{\mu}_{\pm}(\mathbf{K} = \mathbf{0})$. Thus, at the bifurcation point $\tilde{\alpha}_0$ we have

$$\mu_0^- = \frac{\tilde{\alpha}_0^2 - 1}{2} = -\frac{1}{4}, \quad \mu_0^+ = \frac{\tilde{\alpha}_0^2 + 1}{2} = \frac{3}{4}. \tag{A13}$$

We also mention that hereafter, in the arguments of the slowly varying functions, we indicate only the fastest variables, e.g. $A(x_4, y_2)$ means $A(x_4, x_5, \ldots; y_2, y_3, \ldots)$. At the first order in $\epsilon$ Eq. (A2) is satisfied due to Eq. (A4). At the order $\epsilon^{3/2}$ we have:

$$\mu_1 A\boldsymbol{\psi}_0^- + \mu_0^-\boldsymbol{\psi}_1 = -\frac{\partial A}{\partial x_1}\frac{\partial \boldsymbol{\psi}_0^-}{\partial x_0} - i\tilde{\alpha}\sigma_x\frac{\partial A}{\partial x_1}\boldsymbol{\psi}_0^- + H_0\boldsymbol{\psi}_1. \tag{A14}$$

At the bifurcation point $\tilde{\alpha}_0$ we have $\partial\boldsymbol{\psi}_0^-/\partial x_0 \equiv 0$ and $\partial\boldsymbol{\psi}_0^-/\partial y_0 \equiv 0$. Then, taking into account that

$$\boldsymbol{\psi}_0^+ = \sigma_x\boldsymbol{\psi}_0^-, \quad \boldsymbol{\psi}_0^- = \sigma_x\boldsymbol{\psi}_0^+, \tag{A15}$$

we search for the solutions of Eq. (A14) in the form

$$\boldsymbol{\psi}_1 = c\frac{\partial A}{\partial x_1}\boldsymbol{\psi}_0^+, \tag{A16}$$

where $c$ is a constant, resulting for $\mu_0^+ - \mu_0^- = 1$ and $\tilde{\alpha}_0 = 2^{-1/2}$ in the relation

$$\boldsymbol{\psi}_1 = \frac{i}{2^{1/2}}\frac{\partial A}{\partial x_1}\boldsymbol{\psi}_0^+, \quad \mu_1 = 0. \tag{A17}$$

Thus, $\boldsymbol{\psi}_1$ also depends only on the slow variables. Respectively, below it is taken into account that

$$\frac{\partial\boldsymbol{\psi}_1}{\partial x_0} \equiv 0, \quad \frac{\partial\boldsymbol{\psi}_1}{\partial y_0} \equiv 0. \tag{A18}$$

At the order $\epsilon^2$ we obtain

$$\mu_2 A\boldsymbol{\psi}_0^- + \mu_0^-\boldsymbol{\psi}_2 = H_0\boldsymbol{\psi}_2 - \frac{1}{2}\frac{\partial^2 A}{\partial x_1^2}\boldsymbol{\psi}_0^- - i\tilde{\alpha}_0\sigma_x\left(\frac{\partial A}{\partial x_2}\boldsymbol{\psi}_0^- + \frac{\partial\boldsymbol{\psi}_1}{\partial x_1}\right) \tag{A19}$$

Taking into account that $\tilde{\alpha}_0^2 = (\mu_0^+ - \mu_0^-)/2$ and using Eqs. (A18) and (A20), this last equation can be rewritten in the form

$$\mu_2 A\boldsymbol{\psi}_0^- + \mu_0^-\boldsymbol{\psi}_2 = H_0\boldsymbol{\psi}_2 - \frac{i}{2^{1/2}}\frac{\partial A}{\partial x_2}\boldsymbol{\psi}_0^+, \tag{A20}$$

that does not contain second derivative $\partial^2 A/\partial x_1^2$. Thus we obtain

$$\boldsymbol{\psi}_2 = \frac{i}{2^{1/2}}\frac{\partial A}{\partial x_2}\boldsymbol{\psi}_0^+, \quad \mu_2 = 0. \tag{A21}$$

At the next order, $\epsilon^{5/2}$, we obtain:

$$\mu_3 A\boldsymbol{\psi}_0^- + \mu_0^-\boldsymbol{\psi}_3 = H_0\boldsymbol{\psi}_3 - \frac{\partial^2 A}{\partial x_1 \partial x_2}\boldsymbol{\psi}_0^- - \frac{1}{2}\frac{\partial^2\boldsymbol{\psi}_1}{\partial x_1^2}$$
$$-\frac{i}{2^{1/2}}\frac{\partial A}{\partial x_3}\boldsymbol{\psi}_0^+ - \frac{i}{2^{1/2}}\sigma_x\left(\frac{\partial\boldsymbol{\psi}_2}{\partial x_1} + \frac{\partial\boldsymbol{\psi}_1}{\partial x_2}\right). \tag{A22}$$

At the point of bifurcation, we have the relation [see Eqs. (A17) and (A21)]:

$$-\frac{\partial^2 A}{\partial x_1 \partial x_2}\boldsymbol{\psi}_0^- - \frac{i}{2^{1/2}}\sigma_x\left(\frac{\partial \boldsymbol{\psi}_2}{\partial x_1} + \frac{\partial \boldsymbol{\psi}_1}{\partial x_2}\right) = 0, \quad (A23)$$

allowing one to compute

$$\boldsymbol{\psi}_3 = \frac{i}{2^{1/2}}\left(\frac{\partial A}{\partial x_3} + \frac{1}{2}\frac{\partial^3 A}{\partial x_1^3}\right)\boldsymbol{\psi}_0^+, \quad \mu_3 = 0. \quad (A24)$$

Finally, at the order $\epsilon^3$ one obtains:

$$\mu_4 A \boldsymbol{\psi}_0^- + \mu_0^- \boldsymbol{\psi}_4 = H_0 \boldsymbol{\psi}_4 - \frac{1}{2}\frac{\partial^2 A}{\partial y_1^2}\boldsymbol{\psi}_0^- - \frac{1}{2}\frac{\partial^2 \boldsymbol{\psi}_2}{\partial x_1^2}$$
$$-\frac{\partial^2 \boldsymbol{\psi}_1}{\partial x_1 \partial x_2} - \frac{\partial^2 \boldsymbol{\psi}_3}{\partial x_0 \partial x_1} \quad (A25)$$
$$-\frac{i}{2^{1/2}}\sigma_x\left(\frac{\partial A}{\partial x_4}\boldsymbol{\psi}_0^+ + \frac{\partial \boldsymbol{\psi}_3}{\partial x_1} + \frac{\partial \boldsymbol{\psi}_2}{\partial x_2} + \frac{\partial \boldsymbol{\psi}_1}{\partial x_3}\right) - |A|^2 A \boldsymbol{\psi}_0^-,$$

where we have used that $(\boldsymbol{\psi}_0^-)^\dagger \boldsymbol{\psi}_0^- = 1$. Similarly to Eq. (A23) we can show that

$$-\frac{\partial^2 A}{\partial x_1 \partial x_3}\boldsymbol{\psi}_0^- - \frac{i}{2^{1/2}}\sigma_x\left(\frac{\partial \boldsymbol{\psi}_3}{\partial x_1} + \frac{\partial \boldsymbol{\psi}_1}{\partial x_3}\right) = 0. \quad (A26)$$

At this order we only need to satisfy the Fredholm alternative, which in our case is simply the orthogonality of the right hand side of Eq. (A25) to $\boldsymbol{\psi}_0^-$. This yields

$$\mu_4 A = -\frac{1}{2}\frac{\partial^2 A}{\partial y_1^2} + \frac{1}{4}\frac{\partial^4 A}{\partial x_1^4} - |A|^2 A, \quad (A27)$$

where we considered a solution independent of the faster variables (i.e., $x_2$ and $x_3$). After restoring the original variables Eq. (A27) is reduced to Eq. (3).

It is interesting to note that from Eqs. (A11) and (A27) it follows that close to the linear limit, the stable soliton is strongly anisotropic with two orthogonal dimensions (width along the $y$- and $x$-axes) scale as $\epsilon^{1/2}$. This anisotropy of the soliton shape is a direct consequence of the highly anisotropic dispersion of $\mu^-$ in Eq. (A6) [respectively Eq. (2)] at the bifurcation point of the spectrum, containing $\sim K_x^4$ and $\sim K_y^2$ terms. The condition of applicability of this highly anisotropic dispersion approximation, $|K_x| \ll 1$, requires substantial extension along the $x$-axis for solution of Eq. (A27) with typical values of $|x_1| \gg 1$. For realistic experimental parameters [43] this condition corresponds to the condensates containing a few hundreds of atoms.